\documentstyle[12pt,a4wide]{article}
\let\ssection=\section
\renewcommand{\section}{\setcounter{equation}{0}\ssection}

\newcommand{\be}{\begin{enumerate}}
\newcommand{\ee}{\end{enumerate}}
\newcommand{\bi}{\begin{itemize}}
\newcommand{\ei}{\end{itemize}}
\newcommand{\beq}{\begin{equation}}
\newcommand{\eeq}{\end{equation}}
\newcommand{\beqa}{\begin{eqnarray}}
\newcommand{\eeqa}{\end{eqnarray}}

\newcommand{\eq}[1]{(\ref{#1})}

\renewcommand{\a}{\alpha}                   
\renewcommand{\b}{\beta}                    
\renewcommand{\d}{\delta}               
\newcommand{\g}{\gamma}
\newcommand{\k}{\kappa}
\renewcommand{\l}{\lambda}

\newcommand{\bra}[1]{\langle #1 |}

\newcommand{\ket}[1]{| #1 \rangle}

\newcommand{\tr}{{\rm\, tr}}

\newcommand\mathC{\mkern1mu\raise2.2pt\hbox{$\scriptscriptstyle|$}
                {\mkern-7mu\rm C}}
\newcommand{\mathR}{{\rm I\! R}}                

\newcommand{\C}{{\tilde{C}}}
\newcommand{\D}{{\cal D}}
\renewcommand{\P}{{\cal P}}
\newcommand{\PV}{{\cal P}({\cal V})}
\newcommand{\UP}{{\cal UP}}
\renewcommand{\H}{{\cal H}}
\newcommand{\V}{{\cal V}}

\newcommand{\h}[1]{({#1}_{t_1},{#1}_{t_2},\ldots,{#1}_{t_n})}

\begin{document}

\begin{titlepage}
\hspace{10truecm} Imperial/TP/95--96/40\\
\hspace*{11,3truecm} gr-qc/9607051
\vspace*{0,5cm}
\begin{center}
   {\Large\bf Symmetry and History Quantum Theory: \\[3mm]An Analogue of
Wigner's Theorem}
\end{center}
\medskip
\begin{center}
                S.~Schreckenberg\footnote{email: stschr@ic.ac.uk}\\[0.5cm]
                Blackett Laboratory\\
                Imperial College\\
                South Kensington\\
                London SW7 2BZ
\end{center}
\medskip
\begin{center} May 1996\\
[0,5cm] {\em To appear in JMP} 
\end{center}
\vspace{3cm}
\begin{abstract}
The basic ingredients of the `consistent histories' approach to
quantum theory are a space $\UP$ of `history propositions' and a space
$\D$ of `decoherence functionals'. In this article we consider such
history quantum theories in the case where  $\UP$ is given by the set of 
projectors $\P(\V)$ on some Hilbert space $\V$. We 
define the  notion of a `physical symmetry of a history quantum
theory' (PSHQT) and specify such  objects  exhaustively with the aid of 
an analogue of Wigner's theorem. In order to prove  this theorem we 
investigate the structure of $\D$, 
define the notion of an `elementary decoherence functional' and show
that each decoherence functional can be expanded as a certain
combination of these functionals. We call two history quantum theories
that are related by a PSHQT `physically equivalent' and show explicitly, in 
the case of history quantum mechanics, how this notion is compatible with
one that has appeared previously.\\


\end{abstract}
\vfill
\end{titlepage}


\section{Introduction}

In this paper we discuss the mathematical aspects of a notion of
`symmetry' in a history 
quantum theory, such as the decoherent histories approach
to quantum theory initiated by Griffiths
\cite{Gri84}, Omn\`{e}s \cite{Omn88a} and Gell--Mann and Hartle
\cite{GH90a}. Given the major importance symmetries  
play in almost every physical theory, one would like to know what
the  counterpart of this concept is in theories who place the emphasis on 
`histories' and `decoherence functionals' rather than propositions and
states at a fixed time point (as is done in standard quantum theory). This 
involves the problem of giving a meaning to the notion of two history 
theories being equivalent; when consistent sets can be called equivalent,
etc.\cite{GH94}. \\

These  
matters are not settled yet, but I will show that, in case we adopt a
particular notion of `symmetry', it is possible to assign a
well defined meaning to such concepts. In order to understand where these
ideas fit into the structure of such history quantum theories, we have 
to rewrite the decoherent histories approach in a way that
describes the ingredients of such a
theory in a more transparent way. \\

The clarification of the structural
content of  history quantum theories (HQT) is due to 
C.J.~Isham \cite{I94a}, who extracted the basic features of these theories 
in form of a set of axioms which determine the mathematical content of the
framework of such theories. The aim is to place history quantum
theories---as an entirely new approach to the problem of defining and
constructing quantum theories---on an equally firm mathematical base as
other, already 
existing approaches to quantum theories. The explanation of why the
axioms take the particular form chosen is a very deep one and is partly
motivated by problems arising in the area of quantum gravity, in
particular the so called `problem of time'; it uses ideas of
`quasi-temporal' logic and much more. These matters have been
discussed at some length by Isham \cite{I94a}, Isham and Linden
\cite{IL94b} and Schreckenberg \cite{S95b} and the reader is referred to 
those sources for a deeper appreciation of `history quantum theories'. \\

In this paper I will adopt a working, practical approach. I will
mainly restrict 
myself to the case of the history version of finite--dimensional
quantum mechanics. This will prove to be an  ideal model 
to illustrate the concept of `symmetry' introduced in this article. \\

Our discussion will be based on an investigation of  how the
mathematical structure of history quantum theories suggests a notion
of `physical symmetries of history quantum theories'. It is
striking---and indeed very 
satisfying---that the concept developed here is compatible with a
definition and a result presented by Gell--Mann and Hartle in 
\cite{GH94}, even though this was not the original goal of the
enterprise. I was only {\em after} completing the essential parts of
the arguments 
here, that this relation  was made explicit. This speaks on the one hand
for the physical insight and arguments which led Gell--Mann and
Hartle to the notion of `physical equivalence', and, on the other hand, it 
shows the strength of the mathematical formalism developed by 
Isham \cite{I94a} in  order to capture the main ideas of the decoherent 
histories program in a precise mannner. Because of this relation among the 
results, the physical arguments presented in \cite{GH94} can, to some 
extent, be regarded as {\em physical} arguments in favour of the notion 
developed here and, vice versa, the arguments presented here as a 
precise {\em mathematical} statement about such objects, which 
possess a {\em very} transparent description.\\

We will begin with an introduction to the formalism
introduced in \cite{I94a}, recall the classification theorem for
decoherence functionals proven in \cite{ILS94c} and remind the reader
of the content of Wigner's theorem, which will be central to our
investigation. In section \ref{HQT} we reformulate the standard
requirements for `physical symmetries' given by Wigner in a way that
is more suited to our problem 
in that it avoids some of the interpretative difficulties that arise
when one is trying to 
induce a notion of symmetry in history quantum theories from
symmetries defined at a single time point. We proceed by defining
`physical symmetries of a history quantum theory' (PSHQT) and  show 
that a
paricular subset of PSHQT can be induced by unitary or anti--unitary
operators $\hat{U}$ on $\V$, which I call `homogeneous
symmetries'. We show that these symmetries possess a characterisation
{\em \`{a} la  } Wigner. We investigate the structure of the space $\D$
in some detail to show in section \ref{WT} that, in fact, PSHQT
are in one--to--one correspondence with homogeneous symmetries 
and can
thus be characterized by an analogue of Wigner's theorem. We call two 
history quantum theories which are related by a physical symmetry of a
history quantum theory {\em physically equivalent}. This expression  
first appeared in \cite{GH94} and we show
explicitly, for history quantum mechanics, how the notion of
`physical equivalence'---as introduced 
in \cite{GH94}---is naturally induced by a subset of the set of PSHQT. In 
the closing section \ref{SO} we
mention some ways one could try to proceed in order to find a
satisfactory {\em physical}
interpretation of the symmetries considered in this article.  


\subsection{Decoherent Histories and History Quantum Theory}

\subsubsection{The algebraic structure for HQT}
In the decoherent histories approach the two main ingredients are the
so--called `histories', namely sequences of Schr\"odinger picture
projection 
operators $\a:=\h{\a}$, with $t_1<t_2<\cdots<t_n$, defined on the
single time Hilbert space $\H$, and `decoherence functionals', namely
a complex--valued functional $d$ of pairs of  histories. For normal quantum
mechanics the latter is given by:
\beq
        d_{(H,\rho)}(\a,\b):=
            \tr_\H(\C_\a^\dagger\rho_{t_0} \C_\b),      \label{Prob:a1-an}
\eeq
where the `class' operator $\C_\a$ \cite{GH90a} is defined to be 
\beq
\C_\a:=\a_{t_1}(t_1)\a_{t_2}(t_2)\cdots\a_{t_n}(t_n)
\eeq
with
$\{\a_{t_i}(t_i):=
e^{\frac{i}{\hbar}H(t_i-t_0)}\a_{t_i}e^{-\frac{i}{\hbar}H(t_i-t_0)}\}$
being the associated Heisenberg picture operators. For the ease of
exposition, all histories $\a$ will be defined on $n$ arbitrary, but
fixed, time--points.\\

This functional $d(\a,\b)$ is an extension of the formula $d(\a,\a)$
in standard 
quantum mechanics for the  
joint probability of 
finding all the properties  $\a:=\h{\a}$ with $t_1<t_2<\cdots<t_n$ in a
time-ordered sequence of measurements. The aim is to determine with
the aid of certain `consistency conditions' on $d(\a,\b)$ such
histories, on which $d(\a,\a)$ defines a probability distribution.\\  

The expression for the decoherence functional is rather messy: it is
difficult to isolate  the contribution of the histories $\a, \b$ to
the evaluation. Whereas before they were defined as sequences of
Schr\"odinger operators they enter now as a product of Heisenberg
operators. Hence the evolution operator should belong intrinsically to
the decoherence functional as does the density matric $\rho$. \\

The
separation that I have
in mind of the contribution to 
$d_{(H,\rho)}(\a,\b)$ of (i) the histories, and (ii) a part which
encodes all the properties of the decoherence functional,  is best illustrated 
by an analogous expression in standard
quantum mechanics. Namely, 
the probability $p(x\in [a,b])$ of finding that the eigenvalue $x$ of an
observable $X$ lies in the interval $[a,b], a,b\in\mathR $,  given the
state $\rho$ of the system, is evaluated as 
 \beq
p(x\in[a,b],\rho)=\tr_\H( P^X_{[a,b]} \rho ).
\eeq
One can immediately refer to  the density operator to describe the
contribution of the state and to the projection operator
$P^X_{[a,b]}$---as the mathematical representation of the question
asked---to describe 
the contribution of the corresponding observable to the value
$p(x\in[a,b],\rho)$. This is, of course, due to Gleason's theorem which
establishes a one--to--one correspondence between states and density
operators.\\

The appropriate rewriting of the expression for $d_{(H,\rho)}(\a,\b)$,
which is described in  
detail in \cite{ILS94c}, relies on  the
mathematical identity
\beq \label{mi}
\tr_\H (A_1A_2\cdots A_m)=\tr_{\otimes\H^m}(A_1\otimes\cdots\otimes A_m
S)
\eeq
which allows us to express the trace of a product of $m$ operators
$\{A_m\}$ by means of the trace of a single operator
$A_1\otimes\cdots\otimes A_m$ on the $m$--fold tensor product space
$\V_m:=\otimes\H^m$ and a universal operator $S$.\\

Forgetting for a moment the $\rho$ in (\ref{Prob:a1-an}), $d_{(H,\rho)}(\a,\b)$
is given by the product of $2n$ Heisenberg operators. Using formula
(\ref{mi}) we deduce that histories enter the decoherence
functional in the following way:
\beq
\tilde{\a}=\a_{t_1}(t_1)\otimes\a_{t_2}(t_2)\otimes\cdots\otimes\a_{t_n}(t_n)
.
\eeq
Heisenberg operators are just Schr\"odinger operators multiplied on the
left and right by the evolution operators $U(t_i,t_0)$ and its
inverse. Expression (\ref{mi}) shows that this depedence can be thrown
onto the universal operator $S$, allowing us to represent histories by
Schr\"odinger--picture operators 
\beq
\a =\a_{t_1}\otimes\a_{t_2}\otimes\cdots\otimes\a_{t_n}\in
P(\V_n)
\eeq
which contribute to the value for $d_{(h,\rho)}(\a,\b)$ through 
\beq\label{df}
d_{(H,\rho)}(\a,\b)=\tr_{\V_n\otimes\V_n}(\a\otimes\b
X_{(H,\rho)}),
\eeq
for some operator $X_{(H,\rho)}$ defined on
$\V_n\otimes\V_n$. The time--ordered strings of projection operators
$(\a_{t_1},\a_{t_2},\ldots ,\a_{t_n})$ 
are now represented by a homogeneous projection operator
$\a_{t_1}\otimes\a_{t_2}\otimes\cdots\otimes\a_{t_n}$ on the $n$--fold
tensor product space $\V_n=\otimes_{i=1}^n\H_{t_i}$ and one can easily 
see that this association is one--to--one. 
This motivates the definition of a space 
$\UP$ of `history 
propositions' (also called `univeral propositions' or `propositions
about the universe'), which is given by the set of {\em all} projection operators
$\a\in\P(\V_n)$ on the $n$--fold tensor product space . By these means we
have achieved 
the aim of   
separating the contribution of the histories to the value
$d_{(H,\rho)}(\a,\b)$, in that the operator 
$X_{(H.\rho)}$ encodes now all of the 
dynamical information and the initial conditions of the system under
investigation. \\

One can also convince oneself that the decoherence functional
(\ref{df}) satisfies the following properties:   
\begin{eqnarray}\label{r}
&\circ & {\em Hermiticity\/}: d(\a,\b)=d(\b,\a)^\ast\quad\forall
          \a,\b\in \P(\V_n)\\
&\circ & {\em Positivity\/}: d(\a,\a)\ge0\quad\forall \a\in \P(\V_n)\nonumber\\
&\circ & {\em Additivity\/}: d(\a\oplus\b,\g)=d(\a,\g)+d(\b,\g)\nonumber\\
&\circ & {\em Normalisation\/}: d(1,1)=1.\nonumber
\end{eqnarray}
which are the usual
requirements 
for decoherence functionals in the consistent histories approach when
expressed in this formalism. The operation `$\oplus$' is given by the
addition of two 
orthogonal projectors, i.e.\ history propositions, in $\P(\V_n)$. \\

This example seems to suggest that it might be worth trying to define
a {\em history quantum theory} as a theory which has two main ingredients: A
space of history propositions $\UP$ which, in this paper, will be
the space of propositions $\P(\V)$ onto a Hilbert space $\V$;
and a decoherence functional $d\in\D$, where $\D$ 
denotes the space of all decoherence functionals, that is, all those
functionals defined on $\P(\V)\times \P(\V)$ which possess the 
properties (\ref{r}) mentioned
above. Thus $\V$ does not necessarily have to be of the tensor--product
form $\V_n$, and $d\in\D$ will in general not be of the form
$d_{(H,\rho)}$. In this formalism, {\em consistent sets of history
propositions with respect to a $d\in\D$} correspond to certain
partitions of the unit operator on $\V$ into mutually orthogonal
projectors $\{\a_i\}_{i=1}^{m\le\dim\V}$ such that
\beq
d(\a_i,\a_j)=\d_{ij}d(\a_i,\a_i)\quad\forall i,j\in\{1,2,\ldots ,m\}.
\eeq   
The properties (\ref{r}) of $d\in\D$ ensure that the values $d(\a_i,\a_i)$
determine
a probability distribution on the boolean algebra generated 
by the $\{\a_i\}_{i=1}^{m\le\dim\V}$.


\subsubsection{The Classification Theorem}
We want to base our investigations of symmetries on the expression
(\ref{df}) of the decoherence functional. But in order to do so,
we must first be sure that it is not only a lucky coincidence that we are
able to cast this {\em particular} decoherence functional for the
history version of quantum mechanics in the above
form. The formalism for decoherent histories is not simply restricted
to models with unitary evolution, inclusions of final density matrices
and the like. Therefore, in order to formulate a notion of symmetry that is
valid for all these cases, we have to find out whether every decoherence
functional---i.e.~every functional satisfying the properties (\ref{r}) listed
above---can be written in the form (\ref{df}). The clear cut
answer to this is given by the following theorem, see
\cite{ILS94c}, which is valid for any theory in which the history
propositions are given by projectors on a finite--dimensional Hilbert
space $\V$.\\

\noindent
{\bf Theorem}\cite{ILS94c} If $\dim\V>2$, decoherence functionals $d$ are
in one-to-one 
correspondence with operators $X=X_1+iX_2$ on $\V\otimes\V$ according 
to the rule:
\beq
d(\a,\b)=\tr_{\V\otimes\V}(\a\otimes\b X)
\eeq
with the restriction that
\beqa
&a)& X^\dagger=MXM
\quad\mbox{with}\quad\!\!\!M(\ket{v}\otimes\ket{w}):=\ket{w}\otimes\ket{v},
\quad\!\!\!
\forall \ket{v},\ket{w}\in\V.\label{hr}\\
&b)& \tr_{\V\otimes\V}(\a\otimes\a X_1)\ge0 \label{b}\\ 
&c)& \tr_{\V\otimes\V}(X_1)=1.
\eeqa

The restrictions on the operator $X$ on $\V\otimes\V$ reflect the
requirements (\ref{r}). We denote by ${\cal X}_{\cal D}$ the set of
all such operators $X$. This theorem---which has been extended to 
arbitrary von Neumann algebras without factor of
type II in \cite{W95}---is the cornerstone of the
forthcoming investigation. It allows us to shift the investigation of the
properties of decoherence functionals $d\in\D$, where $\D$ denotes the
space of all decoherence functionals, to an analysis of the properties
of the associated operator $X_d\in {\cal X}_{\cal D}$, which, as we 
emphasize once again,
carries all of the `dynamical' content as well as the `initial
conditions' of the model under investigation. \\

It is important to understand the origin of these requirements. 
Condition (\ref{hr}) reflects the hermiticity requirement. The action of $M$
on $\a\otimes\b$ is 
given by $M(\a\otimes\b)M=(\b\otimes\a)$.  Equation (\ref{hr}) follows 
then from the condition 
\[ d(\a,\b)=\tr_{\V\otimes\V}(\a\otimes\b
X)=\tr_{\V\otimes\V}(\b\otimes\a X^\dagger)=d(\b,\a)^\ast.\]
Condition (\ref{b}) stems from the fact that, since $X^\dagger=MXM$ is
equivalent to the pair of conditions  
\beqa
            X_1&=&  MX_1M                         \\    \label{condX1}
            X_2&=& -MX_2M ,                \nonumber         \label{condX2}
\eeqa
it follows that 
\beqa
    \tr_{\V\otimes\V}(\a\otimes\a X_2)&=&-\tr_{\V\otimes\V}
(\a\otimes\a MX_2M)
                                                \\
        &=&-\tr_{\V\otimes\V}\big(M(\a\otimes\a)MX_2\big)=
                        -\tr_{\V\otimes\V}(\a\otimes\a X_2)\nonumber
 \eeqa
 and so $\tr_{\V\otimes\V}(\a\otimes\a X_2)=0$ is implied already
by the hermiticity requirement \eq{hr}.\\

It will be advantageous later to think of $d(\a,\b)$ as the value of
the complex--valued functional
\beqa
\tr : \P(\V)\otimes \P(\V)\times {\cal X}_{\cal D} & \longrightarrow &
\mathC \\ 
 (\a\otimes\b ; X_d) &\mapsto & \tr_{\V\otimes\V}(\a\otimes\b X_d) 
\nonumber
\eeqa


\subsection{Wigner's Theorem}

In order to understand fully the importance of Wigner's result it is
crucial to distinguish between the notion of a {\em symmetry} and that
of a {\em physical symmetry}.\\

\noindent
{\bf Definition} On a complex Hilbert space $\H$ a {\em symmetry} is a
unitary or anti--unitary operator $U$. Thus it leaves invariant the
modulus of the inner product of any pair of two vectors 
$\ket{v}, \ket{w}\in\H$, that is 
\beq
|\langle v, w\rangle |^2 = |\langle Uv, Uw\rangle |^2 ,
\quad\forall \ket{v}, \ket{w}\in\H.
\eeq
This definition is only a mathematical one; it has, {\em a priori},
no motivation by physical arguments. Note also that it 
does {\em not} impose any further defining properties on $U$, such as
commutativity with the 
Hamiltonian operator. Such  requirements only enter at a much later
stage, motivated by analogues of the assumption in classical
mechanics, that a particular physical system evolves along the
flowlines of a specific, Hamiltonian, vectorfield. \\
 
On the other hand, a physical system at a {\em fixed moment of time}
is described in quantum mechanics by a 
state $\sigma\in {\cal S} : \P(\H)\rightarrow\mathR$,
that is, a normalized ($\sigma (1)=1$), positive--valued functional
which is additive on disjoint 
projectors. The states are, via Gleason's theorem, in
one--to--one correspondence with density operators $\rho$ on $\H$
according to the rule
\beq\label{gt}
\sigma(\a,\rho)=\tr_\H(\a\rho ),\quad \forall \a\in \P(\H),
\eeq
where $\rho$ is defined by the properties
\beq
\rho = \rho^\dagger\qquad \rho\ge 0\qquad\tr_\H(\rho)=1.
\eeq
The set of all density operators is often denoted by ${\cal W}_{\cal
S}$.\\

The {\em physical} assumption is now that all that matters are the relations
among the states, which can be entirely described by means of their
overlaps, that is the transition amplitudes
$\tr_\H(\rho_1\rho_2)$. \\

In the finite--dimensional case the self--adjointness of $\rho$
implies that for every density operator there exists an orthonormal
basis $\{\ket{\psi_i}\}$ such that 
\beq
\rho =\sum r_i P_{\ket{\Psi_i}}\qquad\sum r_i =1\qquad r_i\ge 0.
\eeq
Density operators of the form
$P_{\ket{\Psi_i}}:=\ket{\Psi_i}\bra{\Psi_i}\in {\cal W}_{{\cal S}^p}$ are 
called {\em
pure}. Every density operator possesses an expansion in terms of pure
density operators, also representing `rays' of $\H$; instead of ${\cal
W}_{{\cal S}^p}$ we also use sometimes the notation ${\cal R}(\H)$.
Therefore, the  invariance requirement on the transition amplitude 
between two arbitrary 
states $\rho_1,\rho_2$ can be reduced to the requirement that 
\beq
\tr_\H(P_{\ket{\Psi^1}}P_{\ket{\Psi^2}})=\tr_\H(P_{\ket{\Psi^1}}^\xi
P_{\ket{\Psi^2}}^\xi),
\eeq
for arbitrary one--dimensional pure density operators
$P_{\ket{\Psi^1}}, P_{\ket{\Psi^2}}$ and an affine map $\xi:{\cal W}_{{\cal
S}^p} \rightarrow {\cal W}_{{\cal S}^p}$, that is a map satisfying
$\xi(\sum_i c_i P_{\ket{\Psi^i}})=
\sum_i c_i \xi(P_{\ket{\Psi^i}})\equiv\sum_i c_iP_{\ket{\Psi^i}}^\xi ,
c_i\in\mathC$.  \\

\noindent
{\bf Definition} A {\em physical symmetry} is an affine  
bijection $\xi: {\cal W}_{{\cal
S}^p} \rightarrow {\cal W}_{{\cal S}^p} ; P_{\ket{\Psi}}\mapsto
P_{\ket{\Psi}}^\xi$, such that the transition amplitude between pure
density operators remains 
invariant, i.e. that 
\beq
\tr_\H(P_{\ket{\Psi_1}}P_{\ket{\Psi_2}})=
\tr_\H(P^\xi_{\ket{\Psi_1}}P^\xi_{\ket{\Psi_2}}) ,\quad \forall 
\ket{\Psi_1}, \ket{\Psi_2} \in\H .
\eeq
\smallskip

\noindent
{\bf Theorem}\cite{V68} ({\em Wigner}) Every symmetry induces a physical 
symmetry and, conversely,
every one--to--one map $\xi : {\cal W}_{{\cal
S}^p} \rightarrow {\cal W}_{{\cal S}^p} $
preserving orthogonality between rays is a physical symmetry and can
be implemented by a unitary or anti--unitary oprator $U$ on $\H$. 


\section{Symmetry and History Quantum Theory}\label{HQT}

The notion of symmetry discussed in the last section arose from 
discussing quantum mechanics at a single, fixed time point $t\in\mathR$
with a corresponding Hilbert space $\H$. At a single time
point physics is described in terms of the pair $({\cal S},{\cal L})$,
where ${\cal S}$ is the set of states  and ${\cal L}$ is the
lattice of projection operators on $\H$. In order to define physical
symmetries in quantum 
mechanics---in the sense specified by Wigner---only the knowledge of
one part of this 
pair was required, namely the knowledge of the
properties of the set of states ${\cal S}$,  via the map
\beqa\label{q}
\tr : \quad {\cal W}_{{\cal S}^p}\times {\cal W}_{{\cal S}^p}\quad\!\!\!& 
\longrightarrow & \mathR \\
(P_{\ket{\Psi_1}} ; P_{\ket{\Psi_2}}) &\mapsto & 
\tr_\H(P_{\ket{\Psi_1}}P_{\ket{\Psi_2}}).\nonumber
\eeqa
The properties of the
lattice of propositions ${\cal L}=\P(\H)$ did not enter in full. \\

In order to arrive at  a notion of symmetry for HQTs, recall that in a
history quantum theory the pair $(\UP,\D)$ can be seen as a formal  
analogue
of the pair $({\cal L},{\cal S})$.  
Comparing  the two expressions
\beq
\tr_\H(\rho_1\rho_2)\quad\mbox{and}\quad \tr_{\V\otimes\V}(\a\otimes\b X_d)
\eeq
reveals immediately the mathematical difference between them.
In contrast to the quantum mechanical case at a single time point, in HQTs 
the map
\beqa
\tr :\quad \P(\V)\otimes \P(\V)\times {\cal X}_{\cal D} & \longrightarrow & 
\mathC \\
 (\a\otimes\b) \times X_d &\mapsto & \tr_{\V\otimes\V}(\a\otimes\b X_d) 
\nonumber
\eeqa
intertwines the properties of $\UP=\P(\V)$ and ${\cal X}_\D$. This is
not surprising, since the classification theorem is more to be
regarded as an analogue of Gleason's theorem, which in a similar manner
intertwines properties of $\cal L$ and $\cal S$. \\

The invariance 
requirement for the expression $\tr_\H(\rho_1\rho_2)$ has a direct
physical meaning. In HQTs the formal analogue of a density operator
$\rho$ is an operator $X_d\in {\cal X}_\D$ so that the first guess for
symmetries 
might be to look for transformations which leave `transition
amplitudes between different $X_d$' invariant. But such a requirement
would be hard to interpret since HQT deal with `history propositions'
as entities in their own right. The theory is completely 
specified by choosing {\em a particular} decoherence functional, which
is kept fixed throughout. Since the notion of `time' in a specific
history quantum theory is determined by the choice of the structure of the
space of history propositions---for example, the nature of `time' as a
parameter $t\in\mathR$ in quantum mechanics is mirrored in the definition 
of $\P(\V_n)=\P(\otimes_{i=1}^n\H_{t_i})$---and decoherence functionals
associate numbers with these pairs of history propositions {\em as an
entity}, a 
change of the decoherence functional must not occur.\\

How can we nonetheless use, at least at the mathematical level, the 
existing
notion of a physical symmetry and later on Wigner's theorem, to define
a corresponding notion for HQT that does not suffer from the
difficulty mentioned above? The main idea is to characterize the notion of
a {\em physical symmetry} in a form which exploits the pairing
(\ref{gt}) between 
density operators $\rho\in {\cal W}_{\cal S}$ and propositions $\a\in
\P(\H)$ given by Gleason's theorem.


\subsection{Alternative Specification of Physical Symmetries}

We start  by neglecting entirely  the considerations from  which the
expression 
\beq
\tr_\H(P_{\ket{\Psi}}P_{\ket{\Phi}})
\eeq
originally arose. Pure density operators belong trivially to the space of
projection operators $\P(\H)$ and therefore, instead of thinking of the
map (\ref{q})  as
a pairing between states one can think of it as a map
\beqa
\tr : \{{\cal R}(\H)\subset \P(\H)\}\times {\cal W}_{{\cal S}^p} 
\quad\!\!\!&\longrightarrow&\quad\!\!\! \mathR\\
P_{\ket{\Psi}}\times P_{\ket{\Phi}} &\mapsto&
\tr_\H(P_{\ket{\Psi}}P_{\ket{\Phi}})\nonumber
\eeqa
that establishes a pairing between a subset of the space of propositions
and the set 
of pure states.  Therefore, we see immediately that Wigner's result
can be read as
follows: 
Wigner's theorem determines all bijections 
\beqa
\xi : {\cal
R}(\H)\times{\cal W}_{{\cal S}^p} &\rightarrow& {\cal
R}(\H)\times{\cal W}_{{\cal S}^p}\\
(P_{\ket{\Phi}},P_{\ket{\Psi}}) &\mapsto&
(P_{\ket{\Phi}}^\xi,P_{\ket{\Psi}}^\xi),\nonumber
\eeqa
that leave invariant the pairing 
\noindent
\beq
\tr_\H(P_{\ket{\Psi}}P_{\ket{\Phi}})=\tr_\H(P^\xi_{\ket{\Psi}}P^\xi_{\ket{\Phi}}).
\eeq
Now, when seen from this
perspective it is natural to ask whether or not  Wigner's theorem specifies
completely all affine one--to--one maps
\beqa
V : \P(\H)\times {\cal W}_{\cal S} &\rightarrow&
 \P(\H)\times {\cal W}_{\cal S} \label{qm1}\\
 (\a , \rho) &\mapsto& (\a^V , \rho^V)\nonumber 
\eeqa
such that the pairing between propositions und density operators is
left invariant for all 
$\a\in \P(\H)$ and all $\rho\in {\cal W}_{\cal S}$, i.e.
\beq
\tr_\H(\a \rho)=\tr_\H(\a^V\rho^V). \label{qm2}
\eeq
The map is required to be affine since the space ${\cal W}_{\cal S}$
is a convex space. Convex combinations of elements of ${\cal W}_{\cal
S}$ are again density operators.\\

Note that the question posed  is  not trivial: The space of projectors
$\P(\H)$ is a disjoint union of compact Grassmann manifolds and therefore 
allows for a much wider
class of transformation than just unitary or anti--unitary operators
$U$ on $\H$. The three conditions these maps have to satisfy are:
\beqa
&\ast& V: \P(\H) \rightarrow \P(\H) \label{I}\\
&\ast& V: {\cal W}_{\cal S}\rightarrow {\cal W}_{\cal S} \label{II}\\
&\ast& \tr_\H(\a\rho)=\tr_\H(\a^V\rho^V) \label{III}
\eeqa
If we consider transformations on $P(\H)$, the interesting
transformations are given by transforming projectors of different
dimensions to each other. So consider, for example, the
transformation:
\beqa
G:\quad \P(\H) &\rightarrow& \P(\H)\\
\a  &\mapsto& G[\a] \nonumber
\eeqa
whereby  a particular one--dimensional projector $P_{\ket{\Phi}}$ is
mapped into an m--dimensional 
one, $G[P_{\ket{\Phi}}]$, $m>1$. Such a transformation might be 
bijective on $P(\H)$ and even obey
requirement (\ref{III}).\\

Now, regarding $P_{\ket{\Phi}}$ as a pure {\em density operator}, we see 
immediately
that the trace of its image under $G$ is
$\tr_\H(G[P_{\ket{\Phi}}])=m$. Therefore, such a map does not comply
with the requirement (\ref{II}). Thus, only maps which map rays into
rays are allowed and, therefore, all maps obeying the
conditions (\ref{qm1}, \ref{qm2}) are determined by Wigner's
theorem. \\

This reformulation of `physical symmetries'  in terms of the
intersection of different sets of transformations fulfilling
(\ref{I}), (\ref{II}) or (\ref{III}) respectively possesses the advantage
of never having to consider `transition amplitudes between states at a
fixed moment of time'. Physical symmetries just preserve the intertwining
between $({\cal L},{\cal S})$ via Gleason's theorem by transforming
${\cal L}$ {\em and} ${\cal
S}$ by the same tranformation {\em into itself}.   \\

This fact justifies trying to define symmetries of a
history quantum theory by 
exact analogues of the requirements (\ref{I} -- \ref{III}), i.e.\
by replacing   
\beq
\{ \P(\H);{\cal W}_{\cal S};\tr_\H(\a\rho)\} \leftrightarrow \{\P(\V)\otimes
\P(\V); {\cal X}_\D; \tr_{\V\otimes\V}(\a\otimes\b X_d)\}.
\eeq
This notion of a `symmetry of a history quantum theory' does not
suffer from the interpretative difficulty mentioned above.\\

Through this choice, we will build in an invariance requirement for
the values $d(\a,\b)$ of the decoherence functional from the very
start. Some physical arguments for such a choice can be found in
\cite{GH94,DK94a,DK94b}. I will discuss its relevance at various stages in
this paper.


\subsection{Definition and Proposition}

\noindent
{\bf Definition} A{\em physical symmetry of a history quantum theory} 
(PSHQT) is
any affine one--to--one
map  
\beqa
{\cal Q}:{\cal P}(\V)\otimes{\cal
P}(\V)\times {\cal X}_\D &\rightarrow& {\cal P}(\V)\otimes{\cal
P}(\V)\times {\cal X}_\D \\
([ \a\otimes\b ], X_d) &\mapsto& ([\a\otimes\b]^{\cal Q}, X_d^{\cal Q})
\nonumber
\eeqa
that preserves the value of the pairing between history propositions
and operators associated with decoherence functionals, i.e.
\beq\label{cq}
\tr_{\V\otimes\V}(\a\otimes\b
X_d)=\tr_{\V\otimes\V}([\a\otimes\b]^{\cal Q}X_d^{\cal Q}).
\eeq
We state once again the three requirements, such a map has to fulfill:
\beqa
&\ast& {\cal Q}: \P(\V)\otimes\P(\V) \rightarrow 
\P(\V)\otimes \P(\V) \label{Ia}\\
&\ast& {\cal Q}: \qquad\qquad\quad\!\!\! {\cal X}_{\cal D}\rightarrow 
{\cal X}_{\cal D} \label{IIa}\\
&\ast& \hspace{2mm}  \tr_{\V\otimes\V}(\a\otimes\b X_d)=
\tr_{\V\otimes\V}([\a\otimes\b]^{\cal Q}X_d^{\cal Q}) \label{IIIa}
\eeqa
Each condition separately determines a set of
transformations, but only 
the intersection of these sets may be called a PSHQT. The word {\em
physical} is chosen since this definition parallels the one for
physical symmetries given by Wigner. Again, the history propositions
$\a\in\P(\V)$ {\em and} the decoherence functionals, represented by
$X_d\in {\cal X}_\D$, are  transformed {\em together} by the same
transformation. We call two history quantum theories that are related
by a physical symmetry of 
a history quantum theory {\em physically equivalent}. Furthermore, as
will be shown later, this 
definition encompasses the notion of `physical equivalence', first introduced
and justified through physical arguments by
Gell--Mann and Hartle in \cite{GH94}.  It is easy to see that the
following Lemma holds.\\

\noindent
{\bf Lemma} The relation among two history quantum theories
$(hqt_1,hqt_2)$ of being {\em 
physically equivalent}, denoted by $hqt_1 \sim hqt_2$, is an
equivalence--relation. Thus it is (i) reflexive: $hqt_1 \sim hqt_1$,
(ii) symmetric: $hqt_1 \sim hqt_2 \Rightarrow hqt_2 \sim hqt_1$ and
(iii) transitive: $(hqt_1 \sim hqt_2)\, \mbox{and}\, (hqt_2 \sim hqt_3)
\Rightarrow (hqt_1 \sim hqt_3)$. \\ 

There is an obvious class of transformations on $\V\otimes\V$ that
fulfills all three conditions (\ref{Ia} -- \ref{IIa}) stated above:\\

\noindent
{\bf Definition} A {\em homogeneous symmetry} on $\V\otimes\V$ is a
unitary operator $\hat{U}\otimes\hat{U}$ where $\hat{U}$ may be a
unitary or anti--unitary operator on $\V$.\\

\noindent
{\bf Lemma} Every homogeneous symmetry induces a 
PSHQT, i.e.\ $\{HS\}\subset
\{PSHQT\}$.\\

\noindent
{\bf Proof}\\
A homogeneous symmetry induces the maps
\beqa
\a\otimes\b &\mapsto& \hat{U}\otimes\hat{U}[\a\otimes\b
]\hat{U}^\dagger\otimes\hat{U}^\dagger\\
X_d &\mapsto&
\hat{U}\otimes\hat{U}X_d\hat{U}^\dagger\otimes\hat{U}^\dagger
\nonumber
\eeqa
so that for all $\a\otimes\b\in \P(\V)\otimes \P(\V)$ and all $X_d\in
{\cal X}_\D$
it holds that 
\beq
d(\a,\b)=\tr_{\V\otimes\V}\big[(\hat{U}\a \hat{U}^\dagger\otimes \hat{U}\b
\hat{U}^\dagger)(\hat{U}\otimes \hat{U} X_d\hat{U}^\dagger\otimes
\hat{U}^\dagger)\big].
\eeq
One can easily check that 
$\hat{U}\otimes \hat{U} X_d\hat{U}^\dagger\otimes
\hat{U}^\dagger$ fulfills the defining properties for an operator
$X_{d^\prime}\in {\cal X}_\D$ given by the classification
theorem.\hfill $\Box$\\

Homogeneous symmetries possess a different characterisation, that can
easily be derived from Wigner's theorem for quantum mechanics. Recall the
definition of the map $M$ used in the classification theorem for
decoherence functionals,
\beqa
M: \quad\V\otimes\V &\rightarrow& \V\otimes\V \\
\ket{u}\otimes\ket{v} &\mapsto& \ket{v}\otimes\ket{u} ,\nonumber
\eeqa
for all $\ket{u},\ket{v} \in \V\otimes\V$. As a result, its action on
projection operators of the form $\a\otimes\b$ is given by 
\beq
M(\a\otimes\b)M=(\b\otimes\a) ,\quad\forall \a\otimes\b\in
P(\V)\otimes P(\V).
\eeq
In particular, this holds true for $\a,\b\in {\cal R}(\V)$, i.e.\
projection operators 
belonging  to the space ${\cal R}(\V)$ of rays of $\V$. \\

Let $\tau =\tau_1\otimes\tau_2$
denote a map 
\beqa
\tau : {\cal R}(\V)\otimes{\cal R}(\V) &\rightarrow& {\cal
R}(\V)\otimes{\cal R}(\V) \\
P_{\ket{\Psi}}\otimes P_{\ket{\Phi}} &\mapsto&
[P_{\ket{\Psi}}\otimes P_{\ket{\Phi}}]^\tau
:=P_{\ket{\Psi}}^{\tau_1}\otimes P_{\ket{\Phi}}^{\tau_2} , \nonumber
\eeqa
where $\tau_1,\tau_2$ denote transformations on the space of pure density
operators.
A map $\tau$ is said to {\em commute} with $M$, if
$(M\circ\tau)[P_{\ket{\Psi}}\otimes P_{\ket{\Phi}}]=(\tau\circ
M)[P_{\ket{\Psi}}\otimes P_{\ket{\Phi}}]$ for all elements
$P_{\ket{\Psi}}\otimes P_{\ket{\Phi}} \in {\cal
R}(\V)\otimes{\cal R}(\V)$, written symbolically as $[\tau,M]=0$.\\
 
\noindent
{\bf Definition} A {\em homogeneous symmetry of a history 
quantum theory}
(HSHQT) is a one-to--one map ${\tau}:{\cal R}(\V)\otimes{\cal
R}(\V) \rightarrow {\cal R}(\V)\otimes{\cal R}(\V)$ that preserves the
transition amplitude between two elements, i.e.
\[ \tr_{\V\otimes\V}([P_{\ket{\Psi_1}}\otimes
P_{\ket{\Phi_1}}][P_{\ket{\Psi_2}}\otimes P_{\ket{\Phi_2}}])
=\tr_{\V\otimes\V}([P_{\ket{\Psi_1}}\otimes 
P_{\ket{\Phi_1}}]^\tau[P_{\ket{\Psi_2}}\otimes P_{\ket{\Phi_2}}]^\tau)
,\]
and commutes with the map
$M$, that is $[\tau ,M]=0$.\\

\noindent
{\bf Proposition} Every homogeneous symmetry induces a HSHQT and, 
conversely, every one--to-one map ${\tau}:{\cal R}(\V)\otimes{\cal
R}(\V) \rightarrow {\cal R}(\V)\otimes{\cal R}(\V)$ that preserves
orthogonality between the rays and commutes with $M$ is a HSQHT and 
can be implemented by a
unitary or anti--unitary operator $\hat{U}\otimes\hat{U}$ on
$\V\otimes\V$. Symbolically,
\[ \{HSHQT\}\cong\{HS\} .\]

\noindent
{\bf Proof}\\
The transition amplitude between two elements $[P_{\ket{\Psi_1}}\otimes
P_{\ket{\Phi_1}}],[P_{\ket{\Psi_2}}\otimes P_{\ket{\Phi_2}}]\in {\cal
R}(\V)\otimes{\cal R}(\V)$ is given by 
\beq
\tr_{\V\otimes\V}([P_{\ket{\Psi_1}}\otimes
P_{\ket{\Phi_1}}][P_{\ket{\Psi_2}}\otimes
P_{\ket{\Phi_2}}])=\tr_\V(P_{\ket{\Psi_1}}P_{\ket{\Psi_2}}) 
\tr_\V(P_{\ket{\Phi_1}}P_{\ket{\Phi_2}}).
\eeq
Therefore, by Wigner's theorem, all transformations preserving
orthogonality and the transition amplitude can be implemented by
operators $\hat{U}\otimes\hat{V}$, where $\hat{U}$ and $\hat{V}$ are
either unitary or anti--unitary operators on $\V$. Requiring these
transformations to commute with $M$ concludes the proof.\hfill
$\Box$\\

{\em Remark}: It is important to understand why only transformations
of the form $\hat{U}\otimes\hat{U}$ are admitted, and not, for
example, operators of the form 
\beq
\sum_i c_i \hat{U}_i\otimes\hat{U}_i,\quad\sum_i c_i=1,\quad
c_i\in\mathR.
\eeq
 The reason is, that,  starting from $d(\a,\b)$ only
those transformations $\tilde{T}: \a\otimes\b\mapsto
\tilde{T}(\a\otimes\b)\tilde{T}^\dagger$ on $\V\otimes\V$ are allowed
for which 
\beq
\tilde{T}(\a\otimes\b)\tilde{T}^\dagger = \a^\prime\otimes\b^\prime .
\eeq
It is possible for these transformations only to write them as
$d(\a,\b)\mapsto d^\prime(\a^\prime,\b^\prime)$.\\

 We have therefore established the following relation among the three
sets of transformations:
\beq \label{re}
\{HSHQT\}\cong\{HS\}\subset\{PSHQT\} .
\eeq 

We argued that PSHQT determined by conditions (\ref{Ia} -- \ref{IIIa})
is an appropriate notion for symmetries of history quantum
theories. What we want to show now is that all PSHQT are given by
homogeneous symmetries of the form $\hat{U}\otimes\hat{U}$. In view of
(\ref{re}) 
it remains to be shown that $\{HS\}\supset\{PSHQT\}$. \\


\subsection{The structure of $\D$}

In order to show
that all PSHQT can be characterized by means of rays in ${\cal
R}(\V)\otimes{\cal R}(\V)$ we have to discuss in more detail the
structure of the space of decoherence functionals. Comparison with the
case in standard quantum mechanics shows that what we now have to 
look for
is a notion of `elementary decoherence functionals', out of which all other
decoherence functionals can be build by a certain superposition. The
requirement $(\ref{IIIa})$ for PSHQT, 
\[ \tr_{\V\otimes\V}(\a\otimes\b X_d)=
\tr_{\V\otimes\V}([\a\otimes\b]^{\cal Q}X_d^{\cal Q}) ,\]
can then be reduced to a
requirement that has to hold only for all elementary decoherence
functionals. 
We start our investigation with the following observation:\\

\noindent
{\bf Lemma} For any finite set $\{d^{(i)}\}_{i=1}^{n}, d^{(i)}\in\D ,$ it
holds that $d:=\sum_i r_i d^{(i)} \,\in\D , r_i\in\mathR,$
provided that:
\beqa\label{wk}
r_i&\in&\mathR\quad\forall  i\in \{1, 2,\ldots , n\} ,\\
\sum_i r_i d^{(i)}(\a,\a)&\ge&0 \quad \forall \a\in\UP ,\nonumber\\
\sum_i r_i&=&1.\nonumber
\eeqa
These conditions reflect the requirements for $d$ of hermiticity, positivity
and normalization. 
We call such superpositions of decoherence functionals a {\em weak
convex combination} of decoherence functionals. All convex
combinations are weak convex combinations but the converse is not
true. For a convex combination it is required that $r_i\ge 0$;  the
second condition in (\ref{wk}) does not imply $r_i\ge 0$. 
It seems natural to look for so called `pure decoherence functionals'
which can not be written as weak convex combination of other decoherence
functionals. An argument first given by N.~Linden \cite{LP} shows
that {\em any} decoherence functional can be written as the 
sum of two other decoherence functionals. Thus there can be no pure
decoherence functionals. Nonetheless, in this context we are only
interested in a convenient expansion of an arbitrary decoherence
functional by what I will call {\em elementary decoherence
functionals}. This will suffice to prove an analogue of Wigner's theorem 
in the
next section. I will show explicitly how these elementary decoherence
functionals reflect Linden's argument. The same notions apply {\em
mutatis mutandis} for the associated operators $X_d\in{\cal X}_\D$.\\

By the classification theorem \cite{ILS94c} we know that for every
decoherence functional $d\in\D$ its associated operator $X_d$ can be
written as a sum of two self--adjoint operators $X_d=X_1 +iX_2$ subject
to the conditions
\beq
X_1=MX_1M ;\quad  X_2=-MX_2M ;\quad
\tr_{\V\otimes\V}(\a\otimes\a X_1)\ge 0 ;\quad \tr_{\V\otimes\V}(X_1)=1.
\eeq

We seek an expansion for the real part $X_1$ and the 
imaginary part $X_2$ of $X_d$ as a weak convex combination of
decoherence functionals $d^e$ 
that can not be written as a weak convex combination.\\

\noindent
{\bf Proposition} For each $X=X_1+iX_2\in{\cal X}_\D$ there exist two ONB
$\{{\ket{e_i}}\},\{{\ket{b_i}}\}$ on $\V$ such that $X$ can be
written as:
\beqa\label{ed}
X&=&\sum_{i,j} \l_{ij}X_1^{(ij)} + i \sum_{l,m} \k_{lm}X_2^{[lm]}
\eeqa
\vspace{-5mm}
where\\[-5mm]
\beqa
& &X_1^{(ij)}=\frac{1}{2}(P_{\ket{e_i}}\otimes
P_{\ket{e_j}}+P_{\ket{e_j}}\otimes P_{\ket{e_i}}); \label{rp}\\
& &\l_{ij}=\l_{ji},\quad
\sum_{i,j}\l_{ij}=1,\quad\sum_{i,j} a_{ii} \l_{ij} a_{jj}\ge
0,\quad \l_{ij}\in\mathR\nonumber
\eeqa
\vspace{-5mm}
and\\[-5mm]
\beqa& &X_2^{[lm]}=\frac{1}{2}(P_{\ket{b_l}}\otimes
P_{\ket{b_m}}-P_{\ket{b_m}}\otimes P_{\ket{b_l}}); \label{ip}\\
& &\k_{lm}=-\k_{ml},\quad \k_{lm}\in\mathR .\nonumber
\eeqa

{\em Remark}: The positivity requirement
$\tr_{\V\otimes\V}(\a\otimes\a X_1)\ge 0$ gives
rise to the condition $\sum_{i,j} a_{ii} \l_{ij} a_{jj}\ge
0$ for an arbitrary projector $\a=\sum_{ij} a_{ij} \ket{e_i}\bra{e_j}$ on $\V$
when expanded in the basis $\{\ket{e_i}\bra{e_j}\}$.\\

\noindent
{\bf Proof}\\
The proof is a constructive one; it follows the proof of the
classification theorem in \cite{ILS94c}.\\

For each $\a\in \P(\V)$ define a function $d_\a(\b): \P(\V)
\rightarrow \mathC$ where $d_\a(\b):=d(\a,\b)$.
Let $\Re
d_\a$ and $\Im d_a$ denote the real and imaginary parts of $d_\a$, so
that
\beq
    d_\a(\b)=\Re d_\a(\b)+i\Im d_\a(\b)
\eeq
 with $\Re d_\a(\b)\in\mathR$ and $\Im d_\a(\b)\in\mathR$. We  will develop
the argument only for the real part $\Re d_\a(\b)$. The biadditivity 
condition on the $d\in\D$ requires that $\Re d_\a(\b_1\oplus\b_2) =
\Re d_\a(\b_1)+\Re d_\a(\b_2)$ for any orthogonal pair of projectors
$\b_1,\b_2$. Since $d$ is assumed to be bounded, the same holds true
for its real part $\Re d_\a$. For any
$r\in\mathR$, the quantity
 \beq
    \kappa_r(\b):=r\dim(\beta)=r\tr(\beta)
\eeq
 is a real additive function of $\b$, and hence so are $\Re
d_\a+\kappa_r$ for any $r\in\mathR$.\\

In \cite{ILS94c} it was shown that there exists for each $\a\in
\P(\V)$ two real numbers $r_\a, \mu_a \in\mathR$ such that there exists
a density operator $\rho_\a^{\Re}$ on $\V$ for which it holds that
\beq
    \Re d_\a(\b)=\tr_\V\big(({1\over\mu_\a}\rho^{\Re}_\a-r_\a)\b\big)=
                                \tr_\V(Y^{\Re}_\a\b)
\eeq 
where $Y^{\Re}_\a:={1\over\mu_\a}\rho^{\Re}_\a-r_\a$. 
Since $\rho_\a^{\Re}$ is a
density operator, there exists an orthonormal basis
$\{\ket{e_i}\}_{i=1}^{{\rm dim}\V}$ and positive numbers $w^i_\a
\in\mathR$  such that $\rho^{\Re}_\a =\sum_i w^i_\a P_{{\ket{e_i}}}$ and
therefore 
\beq \label{db}
Y^{\Re}_\a=\sum_i (\frac{w^i_\a}{\mu_a}-r_\a) P_{{\ket{e_i}}}.
\eeq

The additivity condition
$d(\a_1\oplus\a_2,\b)=d(\a_1,\b)+d(\a_2,\b)$ implies that
\beq
\tr_\V(Y^{\Re}_{\a_1\oplus\a_2}\b)=\tr_\V(Y^{\Re}_{\a_1}\b)+
\tr_\V(Y^{\Re}_{\a_2}\b)
\eeq
 which, since it is true for all $\b\in\PV$ (and hence for all operators
on $\V$), implies that the operator-valued map $\a\mapsto Y_\a^{\Re}$ is 
itself additive in the sense that
\beq
        Y^{\Re}_{\a_1\oplus\a_2}=Y^{\Re}_{\a_1}+Y^{\Re}_{\a_2}  \label{Yadd}
\eeq
 for all disjoint pairs of projectors $\a_1,\a_2$ on the Hilbert space
$\V$. \\


Let $\{|c_i\rangle\}_{i=1}^{\dim\V} $ be a orthonormal basis of $\cal V
$; let $\{\langle c_i|\}_{i=1}^{\dim\V} $ denote its dual basis. 
Let $\{B_{ij}:=\ket{c_i}\bra{c_j} ;\quad i, j = 1, 2,\ldots, N\}$ be a 
vector-space basis for the operators on
$\cal V $, so that the operators $Y_\alpha^{\Re} $ can be expanded 
as $Y_\alpha^{\Re} =\sum_{i, j=1} y_{ij}^{\Re}(\alpha)B_{ij} $. Then 
relation (\ref{Yadd})
shows that the complex expansion coefficients $y_{ij}^{\Re}(\alpha), i, j=1,
2,\ldots , \dim\V$ must satisfy the additivity condition:
\begin{equation}
y_{ij}^{\Re}(\alpha_1\oplus\alpha_2 )=y_{ij}^{\Re}(\alpha_1
)+y_{ij}^{\Re}(\alpha_2 ). 
\end{equation}

Since $Y_\alpha^{\Re} $ is a bounded operator, its expansion coeffient
functions $\alpha\mapsto y_{ij}^{\Re}(\alpha ) \forall\alpha\in \P(\cal V )$
are bounded as well. It was shown \cite{ILS94c} that there exists an 
operators $\Lambda_{ij}^{\Re}$ on $\cal V $ such that:
\begin{equation}
y_{ij}(\alpha )= \tr_{\cal V }(\alpha \Lambda_{ij}^{\Re}),
\end{equation}
and therefore:
\begin{equation}
Y_\alpha =\sum_{i, j=1}^{N} \tr_{\cal V }(\alpha \Lambda_{ij}^{\Re})B_{ij}.
\end{equation}
In particular,
\beq
   {\Re} d(\a,\b)=
\tr_\V\left\{\sum_{i,j}\big(\tr_\V(\a\Lambda_{ij}^{\Re})\big)B_{ij}\b\right\}
                =\sum_{i,j}\tr_\V(\a\Lambda_{ij}^{\Re})\,\tr_\V(B_{ij}\b). 
\label{3dum1}
\eeq

    We define an operator $X^{\Re}$ on $\V\otimes\V$ by
\beq
            X^{\Re}:=\sum_{ij}\Lambda_{ij}^{\Re}\otimes B_{ij}
\eeq
for which it holds that $\Re d(\a,\b)=\tr_{\V\otimes\V}(\a\otimes\b X^{\Re})$.\\

From now on we choose the particular set of $\{\ket{e_i}\}$
of eigenvectors of the operator $\rho^{\Re}_\a$ as an orthonormal
basis for $\V$, i.e.\ $B_{ij}=\ket{e_i}\bra{e_j}$. 
As an operator on $\V$, $\Lambda_{ij}^{\Re}$ possesses an expansion 
\beq
\Lambda_{ij}^{\Re}=\sum_{k,l} \l^{ij}_{kl} B_{kl}, 
\eeq
so that 
\beq
 X^{\Re}=\sum_{i,j,k,l}\lambda^{ij}_{kl} B_{kl}\otimes B_{ij}.
\eeq
However, from equation (\ref{db}) we see that, by using the basis
$\{\ket{e_i}\}$, this sum reduces to  
\beq\label{r1}
 X^{\Re}=\sum_{i,k,l}\lambda^{ii}_{kl} B_{kl}\otimes B_{ii},
\eeq
since only the $B_{ii}=P_{{\ket{e_i}}}$ appear in the expansion of
$Y_\a^{\Re}$.\\

Remember now, that $X^{\Re}$ stands for the real part $X_1$ of $X_d$, the
operator associated with a decoherence functional $d\in\D$. As such it 
has to fulfill that 
\beq
X^{\Re}=MX^{\Re}M ,
\eeq
where $M$ was defined through the action $M(A\otimes B)M = (B\otimes
A)$ for arbitrary operator $A,B$ on $\V$.
This requirement is strong enough to reduce (\ref{r1}) to 
\beq\label{r2}
 X^{\Re}=\sum_{i,k}\lambda^{ii}_{kk} B_{kk}\otimes B_{ii}.
\eeq

Since $B_{ii}=\ket{e_i}\bra{e_i} = P_{\ket{e_i}}$, we see that the
real part $X^{\Re}\equiv X_1$ of the operator $X_d$ associated with a
decoherence functional $d\in\D$ can be written as 
\beq\label{r21}
 X_1\equiv X^{\Re}=\sum_{i,j}\lambda_{ij} P_{\ket{e_i}}\otimes
P_{\ket{e_j}},  
\eeq
where $\lambda_{ij}:=\lambda^{ii}_{jj}$. It is easy to see that these
coefficient must obey 
\beq
\l_{ij}=\l_{ji},\quad
\sum_{i,j}\l_{ij}=1,\quad\sum_{i,j} a_{ii} \l_{ij} a_{jj}\ge
0,
\eeq
which follow from the requirements of hermiticity, normalization and
positivity. The $a_{ii}\in\mathR$ are expansion coefficients of an
arbitrary projector $\a=\sum_{ij} a_{ij} \ket{e_i}\bra{e_j}$ on $\V$
when expanded in the basis $\{\ket{e_i}\bra{e_j}\}$.\\

Note that the operators  $P_{\ket{e_i}}\otimes
P_{\ket{e_j}}$ are not themselves operators associated with
decoherence functionals. They do not  obey the $X^\dagger=MXM$ 
requirement. By
defining 
\beq
X_1^{(ij)}:=\frac{1}{2}(P_{\ket{e_i}}\otimes
P_{\ket{e_j}}+P_{\ket{e_j}}\otimes P_{\ket{e_i}}),
\eeq
we see that $X_1^{(ij)}$ is an operator, that can be associated with a
decoherence functional. This concludes the proof of the proposition
for the real part. \\

By the same procedure we obtain the
expansion (\ref{ip}) for the imaginary part $X_2$ of $X_d$ in terms of
projectors $P_{\ket{b_i}}$ for a different orthonormal basis 
$\{\ket{b_i}\}$. This concludes the proof.\hfill $\Box$\\

Note that the imaginary part $X_2$ in itself is not an operator that
can be 
associated with a $d\in\D$. Thus, we have shown the following Corollary.\\

\noindent
{\bf Corollary} There exists a one--to--one correspondence between
elementary decoherence functionals $d^e\in\D$ and 
operators $X_{d^e}\in {\cal
X}_\D$ which are given by
the following expression: 
\beqa
X_{d^e}^{(ij)[lm]}=X_1^{(ij)} + i \k_{lm}X_2^{[lm]}, \quad \k_{lm}\in\mathR ,
\eeqa
where  the operators $X_1^{(ij)}, X_2^{[lm]}$ are defined as
above. Note that there is no sum over repeated indices.\\

We have thus shown that every decoherence functional can be written as
a weak convex combination of elementary decoherence functionals. I now
want to show why these $X_{d^e}$ must not be called `pure decoherence
functionals'. \\

We show that every elementary decoherence
functional can be written as the sum of two decoherence
functionals as follows. Due to the fact that the imaginary 
part $X_2^\prime$ of
an operator $X^\prime\in {\cal X}_\D$ associated with arbitrary
decoherence functional  $d^\prime\in\D$ can be
added to the operator $X_{d^e}$ associated with an elementary
decoherence functional $d^e\in {\cal X}_\D$
to produce a new decoherence functional, one can calculate that 
\beq
X_{d^e}=\frac{1}{2}(X_{d^e} + iX^\prime_2) + \frac{1}{2}(X_{d^e} -
iX^\prime_2). 
\eeq                      
Both terms $(X_{d^e} + iX^\prime_2)$ and
$(X_{d^e} - iX^\prime_2)$ in this expression are proper decoherence
functionals, even 
though $iX_2^\prime$ in itself is {\em not} a decoherence functional. 
Thus, elementary decoherence functionals are not pure, but they 
still account for the simplest expansion of an arbitrary 
decoherence functional and this is all 
that is needed for the proof of the analogue of Wigner's theorem.\\

We  have now all the tools at  hand to prove the following theorem.


\section{An Analogue of Wigner's Theorem}\label{WT}

Recall that, up to this point, we know about the following relation between
the sets of `homogeneous symmetries of a history quantum theory',
`homogeneous symmetries' and `physical symmetries of a history 
quantum theory':
\beq 
\{HSHQT\}\cong\{HS\}\subset\{PSHQT\} .
\eeq 
We are now going to show that the sets $\{HS\}$ and $\{PSHQT\}$ are 
identical.

\subsection{The Theorem}

\noindent
{\bf Theorem} There exist a one--to--one correspondence between
homogeneous symmetries and physical symmetries of history quantum
theories. Thus each PSHQT is given by an operator
$\hat{U}\otimes\hat{U}$ and induces a HSHQT; conversely,  every
one--to-one map ${\tau}:{\cal R}(\V)\otimes{\cal 
R}(\V) \rightarrow {\cal R}(\V)\otimes{\cal R}(\V)$ that preserves
orthogonality between the rays and commutes with $M$ can be 
implemented by a
unitary operator $\hat{U}\otimes\hat{U}$ on
$\V\otimes\V$, where $\hat{U}$ may be unitary or anti--unitary.\\

\noindent
{\bf Proof}\\
We will first look for one--to--one maps which leave invariant the pairing
between history propositions 
and decoherence functionals   and map the set of 
rays into itself, {\em and then} restrict  those
transformations to homogeneous symmetries via the condition that they
must map ${\cal X}_\D$ into itself.  \\

We first consider the invariance requirement for the values of
$d(\a,\b)$. This has to hold true for {\em all} decoherence
functionals.
In particular, for the functionals $X_1^{(ij)}$ the relevant number is:
\beq\label{ir1}
2\tr_{\V\otimes\V}(\a\otimes \b X_1^{(ij)})=
\tr_\V(\a P_{\ket{e_i}})\tr_\V(\b P_{\ket{e_j}}) + \tr_\V(\a
 P_{\ket{e_j}})\tr_\V(\b P_{\ket{e_i}}),
\eeq
for some ONB $\{\ket{e_i}\}$ of $\V$.\\

Recall that the requirement of the invariance of the imaginary part of
a decoherence functional leads us to consider
\beq
2\tr_{\V\otimes\V}(\a\otimes \b X_2^{[ij]})=
\tr_\V(\a P_{\ket{b_i}})\tr_\V(\b P_{\ket{b_j}}) - \tr_\V(\a
P_{\ket{b_j}})\tr_\V(\b P_{\ket{b_i}}), 
\eeq
for some ONB $\{\ket{b_i}\}$ of $\V$.\\

Therefore, considering those decoherence functionals $X_d$ for which
$\ket{b_i}=\ket{e_i}$ for all $i\in\{1,2,\ldots ,{\rm dim}\V\}$, and
$\l_{ij}=\k_{ij}$ for all  $i,j\in\{1,2,\ldots ,{\rm dim}\V\}$ we see
that the invariance requirement amounts to requiring that 
\beq \label{wc}
\tr_\V(\a P_{\ket{e_i}})\tr_\V(\b P_{\ket{e_j}})
\eeq
should  remain invariant under the appropriate transformations. At first we
take the case when $\a\otimes\b \in {\cal R}(\V)\otimes{\cal
R}(\V)$. By Wigner's theorem the transformations leaving (\ref{wc})
invariant are given by operators $\hat{U}\otimes\hat{V}$ whereby
$\hat{U}$ and $\hat{V}$ are unitary or anti--unitary operators on
$\V$. Even though these transformations leave the value of $d(\a,\b)$
invariant, they do not comply with the condition of mapping the set 
${\cal X}_\D$ into itself. \\

To see this, 
recall that an element $X_d\in {\cal X}_\D$ is required to satisfy the
condition 
$X_d^\dagger=MX_dM$, which is equivalent to 
$X_1^{(ij)}=MX_1^{(ij)}M$ and $X_2^{[ij]}=-MX_2^{[ij]}M$.    
Consider the following particular decoherence functional $X_1^{(ii)}$
under such a mapping: 
\beq
{\cal X}_\D\ni P_{\ket{e_i}}\otimes P_{\ket{e_i}}\mapsto
P_{\ket{e_i}}^{\hat{U}}\otimes P^{\hat{V}}_{\ket{e_i}}.
\eeq
Since 
\beq
M[P_{\ket{e_i}}^{\hat{U}}\otimes P^{\hat{V}}_{\ket{e_i}}]M\neq
P_{\ket{e_i}}^{\hat{U}}\otimes P^{\hat{V}}_{\ket{e_i}},
\eeq
we see that its  image under the map $\hat{U}\otimes\hat{V}$ does not
belong to ${\cal X}_\D$. To comply with this requirement we need to 
require that 
$[\hat{U}\otimes\hat{V},M]=0$, i.e.\ consider only those operators of
the form $\hat{U}\otimes\hat{U}$.\\

What is now left to show  is that there can be no other transformations
obeying conditions (\ref{Ia} --\ref{IIIa}), even if we allow for
arbitrary tranformations $G=(G_0,G_0)$  
\beqa
G: \P(\V)\otimes\P(\V) &\rightarrow&  \P(\V)\otimes\P(\V) \\
\a\otimes\b &\mapsto& \a^{G_0}\otimes\b^{G_0} \nonumber
\eeqa

The argument is much the same as in the standard quantum mechanical
case: consider a transformation $G=(G_0,G_0)$ that maps
a one--dimensional 
projector $\a\in \P(\V)$ into $\a^{G_0}\in \P(\V)$, an m--dimensional
one, $m>1$. Therefore, 
\[ \a\otimes\a \mapsto \a^{G_0}\otimes\a^{G_0}.\]
It is easy to see that there exist a $d\in\D$ such that
$X_{d}=\a\otimes\a$, see also \cite{S95a}. But since
$\tr_{\V\otimes\V}(\a^{G_0}\otimes\a^{G_0})=m^2$, there exists no
decoherence functional $d^{G_0}\in\D$ for which 
$\a^{G_0}\otimes\a^{G_0}$ is
the associated operator $X_{d^{G_0}}$. 
This concludes the proof.\hfill $\Box$


\subsection{Discussion}

The result of the theorem shows that every PSHQT can be
induced by an unitary or anti--unitary operator $\hat{U}$ on 
$\V$ as follows:
\begin{eqnarray}
&\UP&: \hspace{1,5cm}\a\mapsto \tilde{\a}:=\hat{U}\a\hat{U}^\dagger \\
&\D&    : X_d\mapsto 
   X_{\tilde{d}}:=\tilde{X_d}\equiv\hat{U}\otimes\hat{U}
X_d\hat{U}^\dagger\otimes\hat{U}^\dagger .\nonumber 
\end{eqnarray} 
As a consequence of this transformation the invariance  
\beq
d(\a,\b)=\tilde{d}(\tilde{\a},\tilde{\b})
\eeq
for all $d\in\D$ and all $\a ,\b\in P(\V)$ follows by the property of
the trace.\\

By looking at the definition for a PSHQT from which this theorem arose, it
seems rather unnecessary to proceed via the use of elementary 
decoherence
functionals.  We
could have started immediately by looking for all 
transformations obeying condition (\ref{Ia}); then restrict to those
which map ${\cal X}_\D$ into itself. But since it was not known to
which extent 
${\cal X}_\D$ can accomodate more general transformations $G$ on
$\P(\V)$ it seems a sensible  way to follow this  hybrid path. In particular,
we circumvented the problem of specifying all transformations that map
${\cal X}_\D$ into itself.\\

A central requirement in the proof of the theorem was the invariance
of the value $d(\a,\b)\in\mathC$ for  all pairs  $\a,\b \in
\P(\V)$.
A closer inspection reveals that the existence of 
{\em complex}--valued functionals makes it
possible to reduce the invariance requirement to the form 
(\ref{wc}). However, it is neither necessary to consider {\em
complex}--valued functionals nor to require invariance for
{\em all} pairs $(\a,\b)$. 
We can investigate the
possibility of  softening the
invariance requirement to hold only for 
the `diagonal' values of  $d\in\D$, i.e.~$d(\a,\a)$. Then we are led
to the condition that 
\beq
\tr_{\V\otimes\V}(\a\otimes \a X_1^{(ij)})=
\tr_\V(\a P_{\ket{e_i}})\tr_\V(\a P_{\ket{e_j}})
\eeq
has to remain invariant. In this case we see that we will end up with
the same transformations $\hat{U}\otimes\hat{U}$ on $\V\otimes\V$ as
requiring $d(\a,\b)$ to remain invariant for {\em all} pairs $(\a,\b)$. This is 
due to the fact that
restricting to projectors of the form $\a\otimes\a$ can be formulated
with the aid of the same operator $M$ which is used to formulate the 
defining  property $X^\dagger=MXM$ for  decoherence functionals. 
Therefore, the
requirement on the diagonal part only is strong enough to enforce it
onto the value of $d$ on  any pair $(\a,\b)$.\\

The particular feature of physical symmetries of history quantum theories
is their property of being  implemented by a unitary or anti-unitary operator
$\hat{U}$ on $\V$. As a consequence, each partition of the unit
operator in $\V$ into
mutually orthogonal projectors, that is a set of projectors
$\{\a_i\}$, such that 
\beq
\{\a_i\}_{i=1}^{m\le{\rm dim}\V};\quad\oplus_{i=1}^m\a_i=1 
\eeq
is mapped into another partition of unity. In particular, the
cardinality of this set is preserved. Now, much emphasis in the
decoherent histories approach is placed on finding consistent sets of
history propositions with respect to a particular decoherence
functional $d\in\D$. In the formalism used here, consistent sets  
are naturally associated with particular partitions of unity \cite{S95a}, 
namely those, for which it holds that 
\beq
d(\a_i,\a_j)=\d_{ij}d(\a_i,\a_i)\quad\forall i,j\in\{1,2,\ldots ,m\}.
\eeq

We see therefore that a PSHQT will
always map consistent sets into new consistent sets of the {\em same}
cardinality. There has been some discussion \cite{DK94a,DK94b} whether 
or not one should
allow for transformations between consistent sets of different
cardinality. We see that, at least in this context, this possibility
is excluded if one agrees on the definition of PSHQT presented in this
article.\\


\subsection{Physical symmetries of history quantum mechanics}

The main aim of this section is to show that the notion of `physical
equivalence', introduced in \cite{GH94}, is---when expressed in this
formalism---a particular example of a physical symmetry of history
quantum mechanics. It also serves the purpose of providing the
explicit form of the decoherence functional for this theory.\\

Remember that, for standard quantum mechanics when looked at from the
perspective of the history programme, the space of history
propositions $\a\in\UP$ is given by projectors $\a\in
\P(\V_n)=\P(\otimes_{i=1}^n\H_{t_i})$. The particular decoherence
functional is associated with an operator
\beq
X_{(H,\rho_{t_0},\rho_{t_f})}=
\frac{1}{\tr_\H(\rho_{t_0}\rho_{t_f}(t_f))}\tilde{X}_{(H,\rho_{t_0},\rho_{t_f})}
\eeq
on $\V_n\otimes\V_n$, where I also have inserted a final density operator
at time $t_f$. When 
evaluated on 
homogeneous projectors 
$\a_h=\a_{t_1}\otimes\a_{t_2}\otimes\cdots\otimes\a_{t_n}$ the value
of $d_{(H,\rho_{t_0},\rho_{t_f})}(\a_h,\b_h)$ is given
by 
\begin{eqnarray}
   d_{(H,\rho_{t_0},\rho_{t_f})}(\a_h,\b_h)&=& 
\frac{1}{\tr_\H(\rho_{t_0}\rho_{t_f}(t_f))}\tr_{\V\otimes\V}
(\a_h\otimes\b_h \tilde{X}_{(H,\rho_{t_0},\rho_{t_f})})\\
&=&
\frac{1}{\tr_\H(\rho_{t_0}\rho_{t_f}(t_f))}\tr_\H(\C_{\a_h}^\dagger\rho_{t_0}
\C_{\b_h}\rho_{t_f}(t_f)) , \nonumber
\end{eqnarray}
which coincides with the form of the decoherence functional usually
employed in the histories approach. But note---once again---that
$d_{(H,\rho_{t_0},\rho_{t_f})}(\a,\b)$ is defined {\em for all} $\a
,\b\in\P(\V)$. 
By following the procedure outlined in \cite{ILS94c}, one shows 
that the operator $\tilde{X}_{(H,\rho_{t_0},\rho_{t_f})}$ is given by: 
\begin{eqnarray}\label{dhqm}
\tilde{X}_{(H,\rho_{t_0},\rho_{t_f})} &=&
\big[U(t_1,t_0)^\dagger
\rho_{t_0}U(t_1,t_0)\otimes U(t_2,t_1)^\dagger\otimes\cdots\otimes
U(t_{n},t_{n-1})^\dagger\big]\nonumber\\
&\otimes & 
\big[U(t_2,t_1)\otimes U(t_3,t_2)\otimes\cdots\otimes U(t_{n},t_{n-1})\otimes
 U(t_f,t_n)\rho_{t_f} U(t_f,t_n)^\dagger\big]\nonumber\\[2mm]
&\times & \big( R_{(n)}\otimes 1_{t_1}\otimes 1_{t_2}\otimes\cdots\otimes
1_{t_n}\big)\\
&\times &S_{(2n)} \nonumber\\
&\times &\big( R_{(n)}\otimes 1_{t_1}\otimes
1_{t_2}\otimes\cdots\otimes 1_{t_n}\big). \nonumber
\end{eqnarray}
The last three lines involve universal operators $R_{(n)}, S_{(2n)}$
that arise by rewriting products of operators in terms of
tensor--products \cite{ILS94c}. Thus they are system
independent. This operator 
is defined on $\V_n\otimes \V_n$ and encodes the initial and final
density operators as well as the dynamical evolution in form of the
evolution operator $U(t_i,t_{i-1})$. This is the purest description of
the content of the decoherence functional one can write down. It has a
very transparent form.\\

Recall
\cite{GH94} that ``two triples $(\{C_\a\},H,\rho)$ and
$(\{\tilde{C}_\a\},\tilde{H},\tilde{\rho})$ are called `physically
equivalent' if there are fields and conjugate momenta $(\Phi(x),\pi(x))$
and $(\tilde{\Phi}(x),\tilde{\pi}(x))$, respectively, in which the
triples' histories, Hamiltonian, and initial condition take the same form.'' 
As an example, the  explicit transformation ($\a_{t_i}(t_i)\mapsto
V\a_{t_i}(t_i)V^\dagger,H\mapsto VHV^\dagger, \rho\mapsto V\rho
V^\dagger$) for a fixed unitary operator $V$ on $\H$ was
shown to lead to physically equivalent triples. {\em Remark}: in order not 
to use the same symbol twice, I used
the notation $V$ instead of $U$ as in \cite{GH94} for the unitary
operator; in the context of history quantum mechanics $U(t_i,t_{i-1})$
already denotes the evolution operator.\\
 
First, notice that a transformation of the Heisenberg projection operators
\beq
\a_{t_i}(t_i)\mapsto V\a_{t_i}(t_i)V^\dagger ,\quad\forall i\in\{1,2,\ldots ,n\}
\eeq
 is identical to the 
transformation 
\beq
\a_{t_i}(t_i) \mapsto VU(t_i,t_0)V^\dagger [V\a_{t_i}
V^\dagger]VU(t_i,t_0)V^\dagger ,\quad \forall  i\in\{1,2,\ldots ,n\} 
\eeq
that is a pair of transformations
\beq
\a_{t_i}\mapsto V\a_{t_i}V^\dagger \quad\forall i\in\{1,2,\ldots ,n\} ;
\qquad H\mapsto 
VHV^\dagger , 
\eeq
where the $\a_{t_i}$ denote the Schr\"odinger projection
operators. This is important since in the formalism used here  only a string 
$\a_h=\a_{t_1}\otimes\a_{t_2}\otimes\cdots\otimes\a_{t_n}$  of
{\em Schr\"odinger projection operators} correspond to a homogeneous
history proposition 
$\a_h\in\UP=\P(\V_n)$. Now, defining the unitary operator
$\hat{U}_V:=V\otimes V\otimes\cdots\otimes V$, $n$ times, remembering
that $\rho\mapsto V\rho V^\dagger$ and that the decoherence
functional is given by equation (\ref{dhqm}), we see that the effect of this
transformations on $\tilde{X}_{(H,\rho_{t_0},\rho_{t_f})}$ is given by
$\tilde{X}_{(H,\rho_{t_0},\rho_{t_f})}\mapsto
(\hat{U}_V\otimes\hat{U}_V)\tilde{X}_{(H,\rho_{t_0},\rho_{t_f})}
(\hat{U}_V^\dagger\otimes\hat{U}_V^\dagger)$.
Thus, this transformation between physically
equivalent triples can be described as follows:
\bi
\item The transformation of the history propositions $\a_h\in \P(\V_n)$
as well as 
the operator  
$X_{(H,\rho_{t_0},\rho_{t_f})}$ by a unitary $\hat{U}_V\in B(\V_n)$ clearly
leaves  invariant the values of $d(\a_h,\b_h)$ 
since 
\beq
d(\a_h,\b_h)=\tr_{\V\otimes\V}\big[(\hat{U}_V\a_h \hat{U}_V^\dagger
\otimes \hat{U}_V\b_h
\hat{U}_V^\dagger)(\hat{U}_V\otimes\hat{U}_VX_{(H,\rho_{t_o},\rho_{t_f})}
\hat{U}_V^\dagger\otimes \hat{U}_V^\dagger)\big] 
\eeq
which is the definition of physical symmetry of a history quantum theory
for which $\UP=P(\V_n)$.
Note that, in general, not all unitary operators $\hat{U}$ on $\V_n$ need 
to be  of the form
$\hat{U}_V$ for a unitary operator $V$ on the
single-time Hilbert space $\H$.   
\ei


\section{Summary and Outlook}\label{SO}

In this article we proposed a notion of a `homogeneous symmetry' (HS) and
of a `physical symmetry of a history
quantum theory' (PSHQT). We proved an analogue of Wigner's theorem
which states that there exists a one--to--one
correspondence between both HS and PSHQT. It has been
shown that each PSHQT can be 
induced by a unitary or anti--unitary operator $\hat{U}$ on
$\V$. History quantum theories that are 
related by a PSHQT are called `physically equivalent' and we
showed explicitly in the case of history quantum
mechanics how  this notion encompasses the notion of physical
equivalence introduced by Gell-Mann and Hartle in \cite{GH94} in case
one is dealing with a finte--dimensional, single-time Hilbert space
$\H_t$ at a finte number of time-points $(t_1, t_2,\ldots , t_n)$. An
extension to infinite--dimensional $\H_t$ as well as to a continuous
range of time-points is clearly desirable since such spaces occur
naturally in the context of {\em continuous histories} \cite{IL95a,I95a}.\\
  
In this article  we also  
investigated the structure of the space of decoherence functionals; in
particular, we defined the notion of an `elementary decoherence
functional' in terms of which every decoherence functional can be
expanded. We showed that these decoherence functionals are not pure,
an observation 
that aggrees with a result by Linden \cite{LP} that there exist no
pure decoherence functionals. These elementary decoherence functionals
were employed in order to perform  some 
proofs but have {\em never} been assigned any other status than
a technical one. In particular, we never calculated `transition
amplitudes between decoherence functionals', something that 
entirely contradicts the spirit of history quantum theories. Do these
elementary decoherence functionals possess any physical
interpretation? \\

While the definition of symmetry presented here has very convenient
properties, it does not treat consistent sets of history propositions
in any way preferred to other elements $\a\in \P(\V)$. But since these
are the sets one ultimately wants to determine, it is reasonable to
ask for a notion of symmetry which mirrors their
importance. Reflecting a moment about the structure of consistent sets,
one notices that this amounts to ask for an approach which places
its emphasis on  boolean subalgebras of the space $\P(\V)$. Via the use
of the consistency conditions on the values of the decoherence
functional some of these boolean algebras---namely the ones associated
with  consistent
sets---are selected. Within each of these algebras classical reasoning
without running into logical paradoxes is possible, whereas reasoning
about elements belonging to different consistent sets leads, in
general, 
to inconsistencies in the use of the values $d(\a,\a)$ of the decoherece
functional  as probabilities. The theory 
of `boolean manifolds' \cite{HF81}, which seem to be the most
appropriate objects 
to describe  history quantum theories \cite{I96A,S95b}, 
allows to describes the structure of $\P(\V)$
in these terms and therefore, one is led to the problem of defining a
transformation theory on boolean manifolds. This is a task for future
research. \\

By proving a classification theorem for decoherence functionals---
an analogue of Gleason's theorem in the context of history quantum
theories---and the analogue of Wigner's theorem presented here, we
have laid the 
mathematical foundations for an approach to  quantum theory from
the point of view of the history programme.\\ 

In history quantum theories the decoherence functional can be thought
of as providing the `dynamical' content of the theory. In standard 
quantum mechanics when investigated from the point of view of the
history programme this is manifest in that the space of history 
propositions $\UP$ is given by Schr\"odinger picture projection 
operators and thus represents the `kinematical'---as opposed to
`dynamical'---ingredient of the theory.  In contrast, the decoherence
functional (\ref{dhqm}) 
contains the evolution operator and the initial and final density
operators and thus provides the `dynamical' specification of the
model under investigation. In a companion paper \cite{SS96b} we will 
use the
analogue of Wigner's 
Theorem presented in this article to define and to investigate the
properties of `symmetries of decoherence functionals'.


\bigskip
\noindent
{\large\bf Acknowledgements}

\noindent
I would like to thank Dr.~N.~Linden and  Professor Dr.~C.J.~Isham for 
useful comments.


\end{document}